\documentclass[12pt,letterpaper]{article}

\usepackage{verbatim}
\usepackage{float}
\usepackage{booktabs}
\usepackage{amsmath}
\usepackage{amssymb}
\usepackage{graphicx}
\usepackage{setspace}
\usepackage{natbib}
\usepackage{algorithm}
\usepackage{algorithmic}
\usepackage{xcolor, colortbl}

\definecolor{lightgray}{gray}{0.9}

\bibpunct{(}{)}{;}{a}{,}{,}

\newcommand{\bB}{\mathbf{B}}
\newcommand{\bc}{\mathbf{c}}
\newcommand{\bC}{\mathbf{C}}

\newcommand{\bH}{\mathbf{H}}

\newcommand{\bK}{\mathbf{K}}

\newcommand{\bM}{\mathbf{M}}
\newcommand{\bO}{\mathbf{O}}

\newcommand{\bu}{\mathbf{u}}
\newcommand{\bU}{\mathbf{U}}
\newcommand{\bv}{\mathbf{v}}

\newcommand{\bx}{\mathbf{x}}

\newcommand{\by}{\mathbf{y}}
\newcommand{\bY}{\mathbf{Y}}
\newcommand{\bV}{\mathbf{V}}
\newcommand{\bW}{\mathbf{W}}

\newcommand{\bZ}{\mathbf{Z}}

\newcommand{\vect}[1]{\boldsymbol #1}

\newcommand{\vmu}{\vect{\mu}}

\newcommand{\vepsilon}{\vect{\epsilon}}

\newcommand{\vSigma}{\vect{\Sigma}}

\newcommand{\vXi}{\vect{\Xi}}

\newcommand{\gvn}{\,|\,}
\renewcommand{\epsilon}{\varepsilon}

\renewcommand{\tilde}{\widetilde}

\newcommand{\distn}[1]{\mathcal{#1}}
\newcommand{\matlab}{\mathrm{M}\mathrm{{\scriptstyle ATLAB}}}

\begin{document}

\title{Efficient Estimation of State-Space Mixed-Frequency VARs: A Precision-Based Approach\thanks{We would like to thank Gary Koop for many constructive comments and useful suggestions that have
substantially improved a previous version of the paper}
}
\author{Joshua C. C. Chan \\
Purdue University \and Aubrey Poon \\
 Orebro University \and Dan Zhu \\
 Monash University}

\date{December 2021}

\maketitle

\begin{abstract}

\noindent State-space mixed-frequency vector autoregressions 
are now widely used for nowcasting. Despite their popularity, 
estimating such models can be computationally intensive, especially for large
systems with stochastic volatility. To tackle the computational challenges, 
we propose two novel precision-based samplers to draw the missing observations of the low-frequency variables in these models, building on recent advances in the band and sparse matrix algorithms for state-space models. We show via a simulation study 
that the proposed methods are more numerically accurate and computationally efficient 
compared to standard Kalman-filter based methods. We demonstrate how the proposed method can
be applied in two empirical macroeconomic applications: estimating the monthly output gap and 
studying the response of GDP to a monetary policy shock at the monthly frequency. 
Results from these two empirical applications highlight the importance of incorporating
high-frequency indicators in macroeconomic models.

\bigskip


\bigskip


\end{abstract}

\thispagestyle{empty}

\newpage

\section{Introduction}

Mixed frequency vector autoregressions (MF-VARs) have become increasingly
popular in empirical macroeconomics for forecasting and nowcasting.
 This popularity is due to them being more flexible for incorporating timely information over traditional VARs.  A typical application is to nowcast real GDP, which is only available quarterly, 
using higher-frequency macroeconomic variables or financial indicators. 
There are two common approaches to handle mixed-frequency variables: 
the stacked approach and the state-space approach. The stacked MF-VAR approach of \citet{Ghysels16}
estimates the mixed-frequency model at its lowest observed frequency; recent examples using this approach
include \citet{MOS20} and \citet{KMM20}. In contrast, the state-space approach
of \citet{SS15} treats the high-frequency observations of the low-frequency variables as missing values. These missing values are then estimated using standard Kalman filtering and smoothing algorithms.
Recent applications of this approach include \citet{BBJ19} and \citet{KMMP20}.

One key advantage of the state-space approach compared to the stacked approach is that it explicitly models all variables at the highest observed frequency. It allows the researcher to obtain the interpolated (historical) estimates of the low-frequency variables at a higher frequency. For example, using the state-space approach \citet{SS15} provide monthly GDP estimates, which are required as inputs for a variety of applications. In addition, by modeling all variables at the highest observed frequency, the state-space approach delivers more timely nowcasts.

Nevertheless, the main drawback of the state-space approach is that its estimation tends to be 
computationally intensive. This large computational burden is mainly due to the simulation of the
large number of latent states --- i.e., missing observations of the low-frequency variables --- 
using standard filtering and smoothing techniques. This computational burden becomes prohibitive when
the dimension of the VAR becomes very large. This is a key obstacle in practice for using large state-space MF-VARs, despite the increasing popularity of large VARs in general since the seminal works by \citet{BGR10} and \citet{koop13}.

We overcome this computational challenge by proposing
a novel precision-based approach for drawing the missing observations
of the low-frequency variables in the state-space MF-VAR model.
Precision-based sampling approaches for state-space models were first considered
in \citet{CJ09} and \citet{MMP11}, building upon earlier work by \citet{rue01} on Gaussian Markov 
random fields. Due to their ease of implementation and computational efficiency,
these precision-based samplers are increasingly used in a wide range of empirical applications. 
Recent examples include modeling trend inflation \citep{CKP13,CKP16,chan17,Hou20}, estimating the output gap \citep{GC17b,GC17}, macroeconomic forecasting \citep{CP16, CHP20}, modeling time-varying Phillips curve \citep{Fu20}, and fitting various moving average models \citep{chan13, CEK16, DK20, ZCC20} and dynamic factor models \citep{KS19, BK21}. 
 
Our paper extends the precision-based sampling approach to state-space models with 
missing observations. More specifically, we derive the joint distribution of the missing observations 
(of the low-frequency variables) conditional on the high-frequency data and the model parameters.
With the standard assumptions of Gaussian errors, we show that the
conditional distribution of the missing observations is also Gaussian.
A key feature of this conditional distribution is that its precision matrix is block-banded 
--- i.e., it is sparse and its non-zero elements are arranged along a diagonal band.
As such, the precision-based sampler of \citet{CJ09} can be applied to draw the missing observations. 
In particular, this novel approach allows us to draw all the missing
observations in one step, and is especially efficient compared to standard filtering methods when the observation equation has a complex lag structure.

In addition, we allow the user to impose any linear constraints, both exactly and approximately, 
when sampling the missing observations. This feature is crucial in mixed-frequency applications as linear inter-temporal restrictions are typically imposed to map the high-frequency missing observations to match the observed values of the low-frequency variables. Our paper is related to the recent works by \citet{EKMN20} and \citet{HS21}, who also consider a precision-based sampling approach for settings with missing observations. However, they focus on dynamic factor models and the latter does not consider imposing linear constraints. As such, their methods are not directly applicable to state-space mixed-frequency VARs.

We conduct a series of simulated experiments to illustrate the numerical accuracy and 
computational speed of the proposed precision-based approach. In particular, we estimate the state-space MF-VARs using the proposed samplers and standard filtering methods under a variety of settings. We show that
the proposed precision-based approach has two key advantages. First, it is more computationally efficient compared to standard Kalman-filter based methods and it scales well to high-dimensional settings. Second, it often delivers superior accuracy in estimating the missing observations of the low-frequency variables compared to standard filtering methods. As the number of low-frequency variables increases relative to high-frequency variables, the accuracy of standard filtering methods deteriorates significantly in estimating the missing observations due to numerical issues. 

We demonstrate the proposed precision-based approach using two empirical macroeconomic applications. 
In the first application, we consider a large mixed-frequency Bayesian VAR with 
stochastic volatility and adaptive hierarchical priors to generate 
latent monthly estimates of real GDP and the corresponding output gap estimates 
using the framework of \citet{MW20}. More specifically, we estimate 
a 22-variable mixed-frequency Bayesian VAR consisting of 21 monthly macroeconomic
and financial indicators and a quarterly real GDP measure using the proposed precision-based approach.
We find that the monthly estimates of real GDP track all NBER recession dates well and give plausible values during the COVID-19 pandemic. Furthermore, the monthly output gap estimates, in specific periods, 
can differ from the Congressional Budget Office (CBO) quarterly output gap measure. 
A potential explanation for this difference could be the additional information extracted from the
higher frequency monthly financial  indicators. This highlights the importance of incorporating higher frequency indicators when estimating the output gap.

In the second empirical application, we extend the Bayesian Proxy VAR in \citet{CH19} to
a mixed-frequency setting. More specifically, we expand their VAR with only monthly variables 
to include a quarterly real GDP measure. We find that all the impulse responses of the monthly 
endogenous variables to a monetary policy shock, identified via a proxy variable, display
precisely the same dynamics as presented in \citet{CH19}. However, the key difference is that
the response of real GDP to a monetary policy shock appears to be more subdued compared to
the response of industrial production. For example, the posterior median response of industrial 
production falls to a low of $-$0.5\%, while the posterior median response of real GDP never falls 
below $-$0.2\%. This result therefore suggests that the response of industrial production to 
a monetary policy shock might not be a good proxy for the response of the real economy. This again 
highlights the value of mixed-frequency VARs.

The remainder of the paper is organised as follows. Section 2 discusses
the precision-based sampling approach for drawing the missing
observations in a state-space MF-VAR model. Section 3 presents
the results from a simulation study comparing the proposed 
mixed-frequency precision-based samplers against standard Kalman-filter based 
techniques. Section 4 illustrates how the proposed mixed-frequency
precision-based approach can be applied to two popular empirical macroeconomic
applications. Finally, Section 5 concludes.

\section{Mixed-Frequency Precision-Based Samplers}

This section introduces the proposed precision-based samplers for drawing
the missing observations of the low-frequency variables within a state-space MF-VAR. 
More specifically, in the first subsection, we
derive the conditional distribution of the missing observations given
the observed data for the model and provide an efficient algorithm
to draw from this conditional distribution. Next, in the second subsection,
we show how the draws of the conditional distribution of the missing
observations can be constrained to incorporate inter-temporal restrictions.

\subsection{The Conditional Distribution of the Missing Observations}

Following \citet{SS15}, we express the autoregression at the highest observed frequency. More specifically, let $\by_{t}^{o}$ denote the $n_{o}\times 1$ vector of high-frequency variables that are observed and let $\by_{t}^{u}$ represent the $n_{u}\times 1$ vector of low-frequency variables that are unobserved or only partially observed. A standard example frequently employed in the literature is as follows: $\by_{t}^{o}$ consists of $n_{o}$ monthly variables that are observed at every month $t$, whereas $\by_{t}^{u}$ consists of $n_{u}$ quarterly variables at monthly frequency that are only observed every 3 months (or the linear combination of the 3 monthly variables is observed). 

Then, a mixed-frequency vector autoregression (MF-VAR) with $p$ lags for $\mathbf{y}_{t}=(\by_{t}^{o '}, \by_{t}^{u'})'$ of dimension $n=n_{o}+n_{u}$ can be written as:
\begin{equation}\label{eq:var}
	\mathbf{y}_{t} = \mathbf{b}_0 + \mathbf{B}_{1}\mathbf{y}_{t-1} + \mathbf{B}_{2}\mathbf{y}_{t-2}
	+ \cdots + \mathbf{B}_{p}\mathbf{y}_{t-p} + \vepsilon_{t},\quad \vepsilon_{t}\sim \distn{N}(\mathbf{0},\vSigma),
\end{equation}
where $t=p+1,\ldots,T$, $\mathbf{b}_0 $ is an $n\times 1 $ vector of intercepts, $\bB_{1},\ldots, \bB_p$ are the $n\times n$ VAR coefficients and $\vSigma$ is error covariance matrix. In what follows, the analysis is based on the joint distribution of $\by_{p+1},\ldots, \by_T$ conditional on the initial conditions $\by_1,\ldots, \by_p$.  

Below we derive the joint distribution of the unobserved (low-frequency) variables conditional on the observed (high-frequency) variables. Stacking $\bY=(\mathbf{y}_{1}',\ldots,\mathbf{y}'_{T})'$, we can rewrite \eqref{eq:var} as a standard linear regression in matrix form:
\begin{equation}\label{eq:VAR_stacked}
	\mathbf{H}\mathbf{Y} = \mathbf{c} + \vepsilon,\quad \vepsilon\sim \distn{N}(\mathbf{0},\vXi),
\end{equation}
where $\mathbf{c} = \mathbf{1}_{T-p}\otimes\mathbf{b}_0$, $\vepsilon = (\vepsilon_{p+1}', \ldots, \vepsilon_{T}')'$, $\vXi=\mathbf{I}_{T-p}\otimes\vSigma$, and 
\begin{align*}
 \bH = & \left[
\begin{array}{cccccccc}
-\mathbf{B}_{p} & \cdots & -\mathbf{B}_{1} & \mathbf{I}_{n} & \mathbf{O}_{n} & \cdots & 
\cdots & \mathbf{O}_{n} \\
\mathbf{O}_{n} & -\mathbf{B}_{p} & \cdots & -\mathbf{B}_{1} & \mathbf{I}_{n} & \mathbf{O}_{n} &\cdots & \mathbf{O}_{n} \\
\vdots & \ddots    & \ddots &    & \ddots & \ddots & \ddots & \vdots \\
\mathbf{O}_{n} &  \cdots    & \mathbf{O}_{n} & -\mathbf{B}_{p}  & \cdots & -\mathbf{B}_{1} & \mathbf{I}_{n} & \mathbf{O}_{n} \\
\mathbf{O}_{n} & \cdots & \cdots & \mathbb{\mathbf{O}}_{n} & -\mathbf{B}_{p} & \cdots & -\mathbf{B}_{1} & \mathbf{I}_{n}
\end{array}\right].
\end{align*}
In the above expression, $\mathbf{1}_{T-p}$ is a $(T-p)\times1$ column vector of
ones, $\mathbf{I}_{T-p}$ is the identity matrix of dimension $T-p$, and $\mathbf{O}_{n}$
is the $n\times n$ zero matrix. Note that $\bH$ is of dimension $Tn\times (T-p)n$ and is banded --- i.e., it is a sparse matrix whose non-zero elements are arranged along a diagonal band.

Furthermore, one can write $\mathbf{Y}$ as a linear combination of the observed (high-frequency) and unobserved (low-frequency) variables as:
\begin{equation} \label{eq:M}
	\mathbf{Y}=\mathbf{M}_{o}\mathbf{Y}^{o}+\mathbf{M}_{u}\mathbf{Y}^{u},
\end{equation}
where $\mathbf{Y}^{o}=(\by_{1}^{o'},\ldots,\by_{T}^{o'})'$ is a $Tn_{o}\times1$
vector of the observed variables, $\mathbf{Y}^{u}=(\by_{1}^{u'},\ldots,\by_{T'}^{u})'$
is a $Tn_{u}\times1$ vector of the unobserved  variables, and $\mathbf{M}_{o}$ and $\mathbf{M}_{u}$ are, respectively, $Tn\times Tn_{o}$ and $Tn\times Tn_{u}$ selection matrices that have full column rank. Substituting  \eqref{eq:M} into \eqref{eq:VAR_stacked}, we have
\begin{equation}
	\mathbf{H}(\mathbf{M}_{o}\mathbf{Y}^{o} + \mathbf{M}_{u}\mathbf{Y}^{u})
	= \mathbf{c} + \vepsilon, \quad	\vepsilon\sim \distn{N}(\mathbf{0},\vXi).
\end{equation}
Now, conditional on $\mathbf{Y}^{o}$ (and the model parameters $\bB = (\mathbf{b}_0,\bB_1,\ldots, \bB_p)'$ and $\vSigma$), the joint density of $\mathbf{Y}^{u}$ can be expressed as
\begin{align*}
p(\mathbf{Y}^{u}\gvn\mathbf{Y}^{o},\bB,\vSigma) \propto & 
\exp\left\{-\frac{1}{2}\left(\mathbf{H}(\mathbf{M}_{o}\mathbf{Y}^{o} + \mathbf{M}_{u}\mathbf{Y}^{u})-\bc\right)'\vXi^{-1}\left(\mathbf{H}(\mathbf{M}_{o}\mathbf{Y}^{o}+\mathbf{M}_{u}\mathbf{Y}^{u})-\bc\right)\right\} \\
\propto & \exp\left\{-\frac{1}{2}\left[
\mathbf{Y}^{u'}\mathbf{M}_{u}'\mathbf{H}'\vXi^{-1}\mathbf{H}\mathbf{M}_{u}\mathbf{Y}^{u}
 -2\mathbf{Y}^{u'}\mathbf{M}_{u}'\mathbf{H}'\vXi^{-1}(\bc-\mathbf{H}\mathbf{M}\mathbf{Y}^{o})\right]\right\}.
\end{align*}
Let $\mathbf{K}_{\bY^u}=\mathbf{M}_{u}^{'}\mathbf{H}'\vXi^{-1}\mathbf{H}\mathbf{M}_{u}$, which is a $T n_u\times T n_u$ non-singular matrix (as $\bH\bM_u$ has full column rank). Furthermore, let $\vmu_{\mathbf{Y}^{u}} = \mathbf{K}_{\bY^u}^{-1}\left(\mathbf{M}_{u}^{'}\mathbf{H}'\vXi^{-1}(\bc-\mathbf{H}\mathbf{M}_{o}\mathbf{Y}^{o})\right).$ Then, by completing the square in $\bY^u$, one can write the conditional density of $\mathbf{Y}^{u}$ as 
\begin{align*}
	p(\mathbf{Y}^{u}\gvn\mathbf{Y}^{o},\bB,\vSigma) \propto  &
	\exp\left\{-\frac{1}{2}\left(\mathbf{Y}^{u'}\mathbf{K}_{\bY^u}\mathbf{Y}^{u}
	 -2\mathbf{Y}^{u'}\bK_{\bY^u}\vmu_{\bY^{u}}\right)\right\} \\
	\propto & \exp\left\{-\frac{1}{2}
	\left(\mathbf{Y}^{u}-\vmu_{\mathbf{Y}^{u}}\right)'\bK_{\bY^u}
	\left(\mathbf{Y}^{u}-\vmu_{\mathbf{Y}^{u}}\right)\right\}.
\end{align*}
Thus, we have shown that the conditional distribution of the missing observations given the observed data is Gaussian:
\begin{equation} \label{eq:Yu_cond}
	(\mathbf{Y}^{u}\gvn \mathbf{Y}^{o},\bB,\vSigma )\sim 
	\distn{N}\left(\vmu_{\mathbf{Y}^{u}},\bK_{\bY^u}^{-1}\right).
\end{equation}

Since $\mathbf{H}, \vXi$ and $\bM_u$ are all band matrices, so is the precision matrix 
$\bK_{\bY^u}$. Therefore, we can use the precision-based sampler of \citet{CJ09} to draw 
$\bY^{u}$ efficiently. We also note that the conditional distribution derived in \eqref{eq:Yu_cond} has the same structure even when we allow for time-varying covariance matrices in the state-space MF-VAR model. The only minor change in the expression is that the block-diagonal matrix $\vXi$ now depends on the time-varying covariance matrices $\vSigma_{t}, t=p+1,\ldots, T$.

\subsection{Inter-Temporal Restrictions}

So far the vector of missing observations $\mathbf{Y}^{u}$ is unrestricted. In practice, however, inter-temporal constraints on $\mathbf{Y}^{u}$ are often imposed to map the missing values to the observed values of the low-frequency variables. For example, a commonly employed inter-temporal constraint for log-differenced variables is the log-linear approximation of \citet{MM03, MM10}. More specifically, suppose $y_{i,t}^{u}$ is the missing monthly value of the $i$-th variable at month $t$. Let $\tilde{y}_{i,t}^{u}$ denote the corresponding observed quarterly value (note that $\tilde{y}_{i,t}^{u}$ is only observed for every third month). Then, a standard log-linear approximation to an arithmetic average of the quarterly variable can be expressed as:
\begin{equation} \label{eq:aggre}
	\tilde{y}_{i,t}^{u} = \frac{1}{3}y_{i,t}^{u}+\frac{2}{3}y_{i,t-1}^{u} + y_{i,t-2}^{u}
	+ \frac{2}{3}y_{i,t-3}^{u}+\frac{1}{3}y_{i,t-4}^{u}.
\end{equation}
Stacking the inter-temporal constraints in \eqref{eq:aggre} over time, we obtain
\begin{equation} \label{eq:aggre_stack}
	\tilde{\mathbf{Y}}^{u} = \mathbf{M}_{a}\mathbf{Y}^{u},
\end{equation}
where $\mathbf{M}_{a}$ is the $k\times T n_u$  matrix specifying the $k$ linear restrictions in \eqref{eq:aggre}, and $\tilde{\bY}^{u}$ contains the observed values of the low-frequency variables. As an example, for balanced monthly and quarterly variables, $k = T n_u/3$.

Since for many commonly employed inter-temporal restrictions, the relationships between the missing and observed data are approximate rather than exact, we also consider a version with measurement or approximation errors:
\begin{equation} \label{eq:aggre2_stack}
	\tilde{\mathbf{Y}}^{u} = \mathbf{M}_{a}\mathbf{Y}^{u} + \bu, \quad \bu\sim\distn{N}(\mathbf{0},\bO),
\end{equation}
where $\bO$ is a fixed diagonal covariance matrix that encodes the magnitude of the measurement errors. 

Next, we discuss how one can impose the hard and soft inter-temporal restrictions in \eqref{eq:aggre_stack} and \eqref{eq:aggre2_stack}, respectively. First, to incorporate the information encoded in the hard inter-temporal restrictions, we aim to sample $\mathbf{Y}^{u}$ from the Gaussian distribution in \eqref{eq:Yu_cond} subject to the linear constraint given in~\eqref{eq:aggre_stack}. An efficient way to do so is given in Algorithm 2.6 in \citet{RH05} and Algorithm~2 in \citet{CCZ17}. More specifically, we first draw $\bZ$ from the unconstrained distribution as $\bZ \sim \distn{N}\left(\vmu_{\mathbf{Y}^{u}},\bK_{\bY^u}^{-1}\right)$. We then correct for the constraint by computing 
\[
	\bY^{u} = \bZ + \bK_{\bY^u}^{-1} \mathbf{M}_{a}^{'}(\mathbf{M}_{a}\bK_{\bY^u}^{-1}\mathbf{M}_{a}^{'})^{-1}(\tilde{\bY}^{u} -\mathbf{M}_{a}\bZ).
\]
It can be shown that $\bY^{u}$ has the correct distribution, i.e., it follows the $\distn{N}\left(\vmu_{\mathbf{Y}^{u}},\bK_{\bY^u}^{-1}\right)$ distribution satisfying the constraint 
$\tilde{\bY}^{u} = \mathbf{M}_{a}\bY^{u} $. Algorithm~\ref{alg:gaulr} describes an efficient implementation in \citet{RH05} that avoids explicitly computing the inverse of $\bK_{\bY^u}$ or $\mathbf{M}_{a}\bK_{\bY^u}^{-1}\mathbf{M}_{a}^{'}$. Using this implementation, the additional computational cost for correcting the constraint is relatively low for $k\ll Tn_u$. For large $k$, this algorithm would involve a few large, dense matrices, and the computations could be more intensive. 

\begin{algorithm}[H]
\caption{Sampling $(\mathbf{Y}^{u}\gvn\tilde{\bY}^{u} = \mathbf{M}_{a}\bY^{u})$ with hard inter-temporal restrictions, where $\mathbf{Y}^{u}\sim \distn{N}\left(\vmu_{\mathbf{Y}^{u}},\bK_{\bY^u}^{-1}\right)$.}
\label{alg:gaulr}
Given the parameters $\vmu_{\mathbf{Y}^{u}}$ and $\bK_{\bY^u}$, complete the following steps.
\begin{enumerate}
	\item Obtain the Cholesky factor $\bC$ of $\bK_{\bY^u}$ such that $\bK_{\bY^u}=\bC\bC'$
	
	\item Solve $\bC'\bv=\bx$ for $\bv$ by backward substitution, where $\bx\sim\distn{N}(\mathbf{0},\mathbf{I}_{Tn_u})$
	
	\item Compute $\bZ = \vmu_{\mathbf{Y}^{u}} + \bv$
	
	\item Solve $\bC\bC'\bV = \bM_{a}'$ for $\bV$ by forward and backward substitution
	
	\item Solve $\bW\bU =\bV'$ for $\bU$, where $\bW = \bM_{a}\bV$
		
	\item Return $\mathbf{Y}^{u} = \bZ + \bU'(\tilde{\bY}^{u} - \bM_{a}\bZ)$
	
\end{enumerate}
\end{algorithm}

Next, to incorporate the soft inter-temporal restrictions, one can view \eqref{eq:aggre_stack} as a new measurement equation and the Gaussian distribution of $\mathbf{Y}^{u}$ in \eqref{eq:Yu_cond} as the 
``prior distribution". Then, by standard linear regression results, we obtain
\begin{equation} \label{eq:Yu_cond_sc}
	(\mathbf{Y}^{u} \gvn \mathbf{Y}^{o},\tilde{\bY}^{u}, \bB,\vSigma )\sim 
		\distn{N}\left(\overline{\vmu}_{\mathbf{Y}^{u}},\overline{\bK}_{\bY^u}^{-1}\right),
\end{equation} 
where
\[
	\overline{\bK}_{\bY^u} = \bM_a'\bO^{-1}\bM_a + \bK_{\bY^u}, \quad
	\overline{\vmu}_{\mathbf{Y}^{u}} = \overline{\bK}_{\bY^u}^{-1}\left(\bM_a'\bO^{-1}\tilde{\bY}^{u} 
	+ \bK_{\bY^u}\vmu_{\mathbf{Y}^{u}}\right).
\]
Since the matrices $\bM_a, \bO$ and $\bK_{\bY^u}$ are all banded, so is $\overline{\bK}_{\bY^u}$. Hence, the precision-based sampler of \citet{CJ09} can be directly applied to sample $\mathbf{Y}^{u}$ efficiently. Compared to Algorithm~\ref{alg:gaulr} for the hard inter-temporal constraints, sampling from \eqref{eq:Yu_cond_sc} is much faster and scales well to high-dimensional settings. For approximate inter-temporal restrictions such as \citet{MM03, MM10}, the latter sampler is naturally preferable. For other exact inter-temporal restrictions, empirically one can approximate these hard restrictions by setting the diagonal elements of $\bO$ to be very small (e.g., $10^{-8}$). Therefore, we use the sampling scheme in \eqref{eq:Yu_cond_sc} as the baseline.

\section{A Simulation Study}

We conduct a simulation study to assess the speed and accuracy of the proposed precision-based methods for drawing the latent missing observations of the low-frequency variables relative to Kalman-filter based methods. In what follows, all data-generating processes (DGPs) assume the following VAR structure with $p=4$ lags: 
\[
	\by_{t} = \mathbf{b}_0 + \bB_{1}\by_{t-1} + \bB_{2}\by_{t-2} + \bB_{3}\by_{t-3} + \bB_{4}\by_{t-4} + 
	\vepsilon_{t}, \quad \vepsilon_{t}\sim \distn{N}(\mathbf{0},\vSigma),\label{eq:9}
\]
where $\by_{t}=(\by_{t}^{o'},\by_{t}^{u'})$ is an $n\times1$ vector of mixed-frequency data, 
$\by_{t}^{o}$ is the $n_{o}\times 1$ vector of the high-frequency variables 
and $\by_{t}^{u}$ is the $n_{u}\times1$ vector of the low-frequency variables. 
We set $\mathbf{b}_0 = 0.01\times \mathbf{1}_{n}$, $\mathbf{B}_{1}=0.5\times\mathbf{I}_{n}$,
$\mathbf{B}_{2}=0.05\times\mathbf{I}_{n}$, $\mathbf{B}_{3}=0.001\times\mathbf{I}_{n}$,
$\mathbf{B}_{4}=0.0001\times\mathbf{I}_{n}$, and $\vSigma=0.01\times\mathbf{I}_{n}$.


We consider DGPs of different dimensions with $T=500$: small ($n=5,n_{o}=4,n_{u}=~1)$, medium ($n=11, n_{o}=10, n_{u}=1)$ and large ($n=21, n_{o}=20,n_{u}=1)$. We also investigate settings with a larger number of unobserved variables ($n_{u}=5$). For each simulated dataset $r=1,\ldots, R$, we estimate the missing observations $\by_{t}^{u,r}$ using 3 methods: precision-based sampler with hard inter-temporal constraints in \eqref{eq:aggre_stack}, precision-based sampler with soft constraints in \eqref{eq:aggre2_stack}, and the simulation smoother of \citet{CK94} as implemented in the code provided by \citet{SS15}. Lastly, for all VARs we assume the standard normal-inverse-Wishart prior with non-informative hyperparameters.

To assess the accuracy of the proposed methods, we compute the mean squared errors (MSEs) of the estimated missing observations against the actual simulated values. The MSEs of all methods are computed using $R=10$ simulations, and the results are displayed in Table~\ref{tab:MSE}. We also report the computation times, based on 20,000 MCMC draws with a burn-in period of 10,000 draws\footnote{The computation times are based on a standard desktop with an Intel Core i7-7700 @ 3.6GHz processor and 16 GB of RAM and the code is implemented in $\matlab$.}. Since the three methods aim to draw from the same distribution --- namely, the conditional distribution of the missing observations given the observed data and model parameters --- in principle they should give the same MSEs (modulo Monte Carlo errors). Indeed, they tend to give very similar MSEs when the ratio $n_o/n_u$ is sufficiently large. However, when $n_u$, the number of variables with missing observations, is large relative to $n_o$, the  number of fully observed variables, the accuracy of the Kalman-filter based method appears to deteriorate significantly relative to the proposed methods, possibly due to numerical errors. In terms of runtime, it is clear that the proposed precision-based methods are more computationally efficient compared to the Kalman-filter based method across a range of settings. 

\begin{table}[H]
\caption{Mean squared errors of the estimated missing observations and computation times using three methods: the proposed precision-based method with hard inter-temporal constraints (Precision-hard), the precision-based method with soft inter-temporal constraints (Precision-soft) and the simulation smoother of \citet{CK94} implemented in \citet{SS15} (KF).}
\label{tab:MSE}
\centering
\begin{tabular}{cccccccc} \hline\hline
	&		&	\multicolumn{3}{c}{MSE}					&	\multicolumn{3}{c}{Computation time (minutes)}				\\
$n_u$	&	$n_o$	&	Precision-hard	&	Precision-soft	&	KF &	Precision-hard	&	Precision-soft	&	KF	\\ \hline
1	&	4	&	0.014	&	0.024	&	0.029	&	0.8	&	1	&	6	\\
\rowcolor{lightgray}
1	&	10	&	0.015	&	0.024	&	0.019	&	3	&	3	&	8	\\
1	&	20	&	0.018	&	0.023	&	0.021	&	25	&	26	&	36	\\
\rowcolor{lightgray}
5	&	4	&	0.899	&	0.222	&	2.140	&	13	&	3	&	26	\\
5	&	10	&	0.035	&	0.029	&	0.070	&	11	&	10	&	30	\\
\rowcolor{lightgray}
5	&	20	&	0.023	&	0.025	&	0.027	&	64	&	54	&	73	\\ \hline \hline
\end{tabular}
\end{table}

Table~\ref{tab:MSE} reports the runtimes of full MCMC estimation. When the dimension of the VAR increases, the part of the posterior sampler that simulates the VAR coefficients dominates, and it gives the impression that the runtimes of the three methods converge. To better understand how the proposed methods perform across a wider range of settings, next we compare only the runtimes of sampling the missing observations. 

\begin{figure}[H]
    \centering
   \includegraphics[width=.95\textwidth]{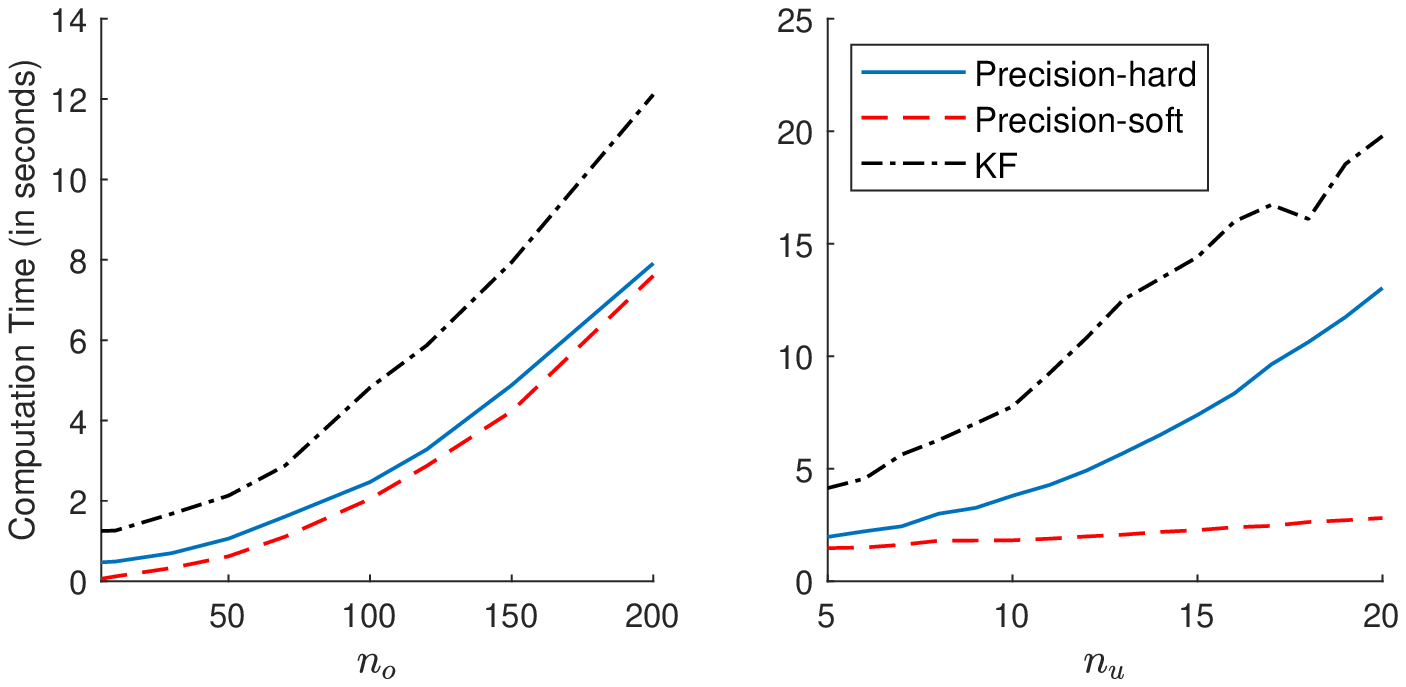}
   \caption{Computation times of obtaining 10 draws againist $n_o$ and $n_u$, the numbers of observed and partially unobserved variables, respectively, with $T=500$ and $p=4$. The three methods are: precision-based sampler with hard inter-temporal constraints (Precision-hard), precision-based sampler with soft constraints (Precision-soft) and the simulation smoother of \citet{CK94} implemented in \citet{SS15} (KF).}
   \label{fig:simulation_n}	
\end{figure}

First, Figure~\ref{fig:simulation_n} reports the runtimes of sampling 10 draws of the missing observations using the three methods for a range of $n_o$ and $n_u$. It is clear that both precision-based methods compare favorably to the Kalman-filter based method, and both scale well to high dimensional settings. In addition, the variant with soft constrictions is especially efficient when there are a large number of variables with missing observations. 

\begin{figure}[H]
    \centering
   \includegraphics[width=.95\textwidth]{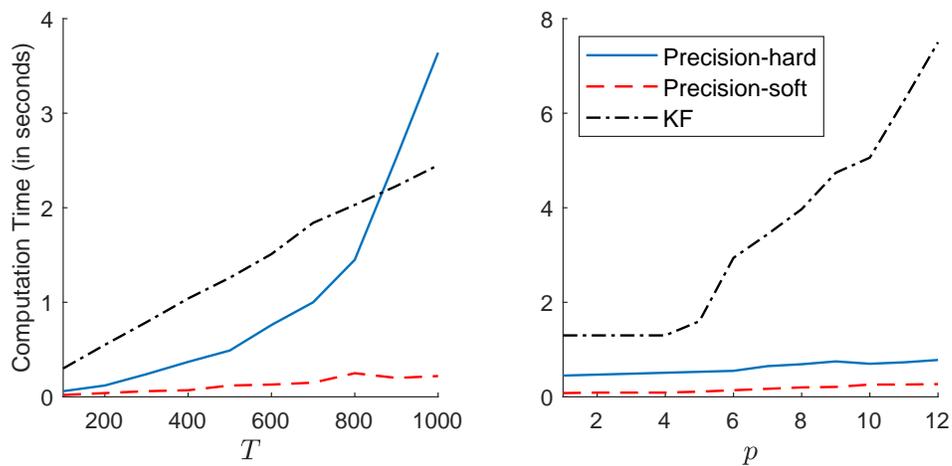}
   \caption{Computation times of obtaining 10 draws against $T$ and $p$, the numbers of time periods and lags, respectively, with $n_u=5$ and $n_o=10$. The three methods are: precision-based sampler with hard inter-temporal constraints (Precision-hard), precision-based sampler with soft constraints (Precision-soft) and the simulation smoother of \citet{CK94} implemented in \citet{SS15} (KF).}
   \label{fig:simulation_pT}	
\end{figure}

Next, Figure~\ref{fig:simulation_pT} reports the runtimes of sampling ten draws of the missing observations for a range of sample sizes $T$ and lag lengths $p$. While both precision-based methods perform well, the version with soft constrictions does substantially better and scales well to very large $T$ and $p$. It is also worth mentioning that to apply the Kalman filter, and one needs to redefine the states so that the observation equation depends only on the current (redefined) state. When $p$ is large, the dimension of this new state vector is large. And that is the reason why the Kalman-filter based method becomes very computationally intensive when $p$ is large. In contrast, the computational costs of the precision-based methods remain low even for long lag lengths.

\section{Empirical Applications}

We demonstrate the proposed mixed-frequency precision-based samplers via two popular empirical macroeconomic applications. First, we show that the proposed samplers can be incorporated efficiently
within a state-of-the-art large Bayesian VAR with stochastic volatility and global-local priors.
Furthermore, we also show that we can derive monthly estimates of the US output gap using the framework of \citet{MW20}. In the second application, we extend the methodology in \citet{CH19} by proposing
a novel mixed-frequency Bayesian Proxy VAR estimated using the proposed samplers. 

\subsection{A State-Space Mixed-Frequency VAR with Stochastic Volatility and Global-Local
Priors}

Since the seminal works by \citet{BGR10} and \citet{koop13}, large Bayesian VARs have become
increasingly popular in empirical macroeconomics. However, large
Bayesian VARs tend to be over-parameterised, and a key way to overcome this
problem is to implement shrinkage priors, such as the Minnesota prior \citep{DLS84, litterman86} 
and the more recent global-local priors \citep[see][]{PS10, HF19, CHP20}\footnote{For a textbook treatment of shrinkage priors for large VARs, see, e.g., \citet{KK10}, \citet{karlsson13} and \citet{chan20_chapter}.}. Another popular feature researchers and practitioners incorporate within a large Bayesian VAR is stochastic volatility (SV). There are a large number of studies that document the importance of allowing for SV (or a time-varying covariance structure) when modelling macroeconomic data \citep[see, e.g.,][]{clark11, CR15, CCM19}.

Recently, \citet{chan21} introduced a class of Minnesota-type adaptive hierarchical priors for large Bayesian VARs with SV. In a nutshell, these new priors combine the advantages of both the Minnesota prior (e.g., rich prior beliefs such as cross-variable shrinkage) and global-local priors (e.g., heavy-tails and substantial mass around 0). These new priors are shown to provide superior forecasts compared to both the Minnesota prior and conventional global-local priors. 

In this first application, we extend the large Bayesian VAR in \citet{chan21} to a mixed-frequency state-space setting and estimate the model using the proposed precision-based samplers. We mostly follow the model assumptions specified in \citet{chan21} and estimate a 22-variable mixed-frequency VAR with SV (denoted here as MF-BVAR-SV) with $p=4$ lags. The variables consist of 21 monthly macroeconomic indicators and a quarterly real GDP measure\footnote{It takes approximately 29 minutes to estimate the MF-BVAR-SV model
for 20,000 MCMC draws using a standard desktop as specified above.}. This exercise is motivated by recent interest in more timely monthly estimates of real GDP \citep[see, e.g.,][]{BBK19}. A more timely measure of GDP allows policymakers to react to economic shocks more rapidly. Therefore, our primary focus of this application is in generating the latent monthly estimates for real GDP and the corresponding output gap. 

The complete list of the 22 variables and their transformations are presented in Table~\ref{tab:data} in the appendix. The sample covers 1960M1-2021M3 (and 1960Q1-2021Q1), which includes the pandemic period. The 21 monthly macroeconomic indicators are selected so that they are broadly similar to the dataset considered in \citet{MW20}. The majority of the variables are log-differenced to ensure stationarity. In addition, we impose
the standard inter-temporal constraint of \citet{MM03,MM10} on the quarterly real GDP
\[
	y_{t}^{GDP,Q}=\frac{1}{3}y_{t}^{GDP,m}+\frac{2}{3}y_{t-1}^{GDP,m}+y_{t-2}^{GDP,m}+\frac{2}{3}y_{t-3}^{GDP,m}+\frac{1}{3}y_{t-4}^{GDP,m},
\]
where $y_{t}^{GDP,Q}$ and $y_{t}^{GDP,m}$ are the observed quarterly and latent monthly real GDP variables at time $t$, respectively.

\begin{figure}[H]
	\centering
	\includegraphics[width=1\textwidth]{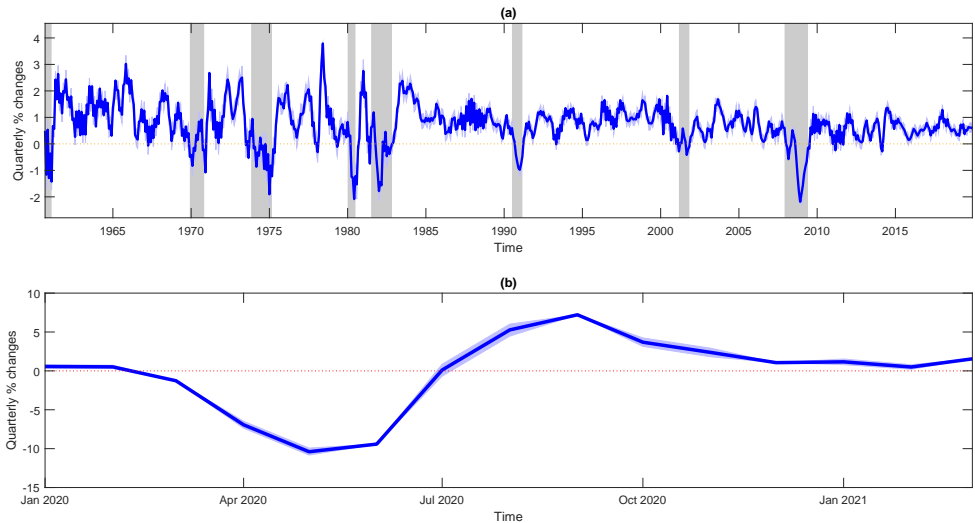}
	\caption{Estimates of monthly Real GDP. Notes: The blue line represents the posterior medians, and the light blue shaded area represents the 68 per cent credible interval. Panel (a) plots the monthly Real GDP estimates for the pre-pandemic period. Panel (b) plots the monthly Real GDP estimates during the pandemic period. The grey shaded areas denote NBER recession dates.}
	\label{fig:GDPprecision}
\end{figure}

\begin{figure}[H]
\centering
\includegraphics[width=1\textwidth]{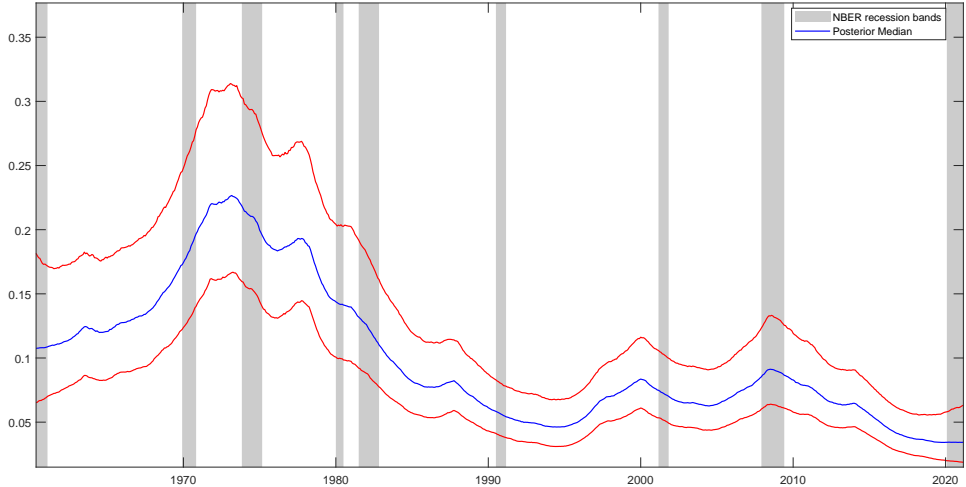}
\caption{Estimates of the stochastic volatility for Real GDP. Notes: The blue line represents the posterior median estimates, and the red line represents the 68 per cent credible interval. The grey shaded areas denote NBER recession dates.}
\label{fig:GDPSV}
\end{figure}

Figure~\ref{fig:GDPprecision} plots the monthly real GDP estimates (posterior median). Panel~(a) plots the monthly real GDP estimates during the pre-pandemic period of 1960-2019, and panel (b) plots the monthly estimates during the pandemic period. There are two key features we can draw from Figure~\ref{fig:GDPSV}. First, our monthly real GDP estimates capture all the relevant NBER recessions during the pre-pandemic period. Second, they are consistent with the observed quarterly values produced during the pandemic. This highlights that the proposed precision-based methods are robust to handle samples with extreme outliers. 

Next, in figure \ref{fig:GDPSV}, we plot the SV estimates of real GDP from the MF-BVAR-SV model. The pattern displayed in these SV estimates is consistent with the historical performance of the US economy. For instance, the high volatility associated with the start of the sample can be attributed to the oil price shocks of the 1970-80s. On the other hand, the decline in volatility from the 1980s to the early 2000s is consistent with the timing of the Great Moderation. Therefore, we conclude that the proposed precision-based samplers work well and can handle samples with extreme outliers.

Another major advantage of producing monthly real GDP estimates from
the MF-BVAR-SV model is that we can derive the corresponding monthly
output gap via a multivariate Beveridge and Nelson (BN) trend-cycle
decomposition from \citet{MW20}. Recently, \citet{BMW21} applied this 
BN trend-cycle decomposition to a mixed-frequency VAR to nowcast the US output gap.
However, their mixed-frequency VAR follows the stacked approach where 
the model is expressed at the lowest observed frequency. More specifically, their mixed-frequency VAR 
is essentially a multi-equation U-MIDAS model. Therefore, it can only 
produce a quarterly measure of the output gap.
In contrast, our MF-BVAR-SV model is a state-space mixed-frequency 
model where we can directly produce a higher-frequency monthly estimate
of the output gap. To the best of our knowledge, this is the first 
study to apply the multivariate BN trend-cycle decomposition 
within a state-space mixed-frequency VAR framework. 

Following \citet{MW20}, a finite-order VAR$(p)$ can be
represented in the following companion form:
\begin{equation}
	(\mathbf{X}_{t}-\vmu)=\mathbf{F}(\mathbf{X}_{t-1}-\vmu)+\mathbf{G}\mathbf{e}_{t},\label{eq:var2}
\end{equation}
where $\mathbf{X}_{t}$ is vector of the $n$ stationary variables, $\vmu$ is a vector of the unconditional means $\mathbf{F}$ is the companion matrix, $\mathbf{G}$ maps the VAR forecast errors to the companion form, and $\mathbf{e}_{t}$ is a vector of forecast errors. Note that the unconditional means can be written as $\vmu=(\mathbf{I}_{n}-\mathbf{B}_{1}-\mathbf{B}_{2}-\cdots-\mathbf{B}_{p})^{-1}\mathbf{b}_0$, where $\mathbf{B}_{1},\ldots, \mathbf{B}_{p}$ are the VAR coefficient matrices and 
$\mathbf{\mathbf{b}_0}$ is the vector of the intercepts.

From \eqref{eq:var2}, we can derive the cyclical component of the multivariate BN trend-cycle
decomposition as
\[
	\bc_{t}^{BN}=-\mathbf{F}(\mathbf{I}-\mathbf{F})^{-1}(\mathbf{X}_{t}-\vmu).
\]

\begin{figure}[H]
\centering
\includegraphics[width=1\textwidth]{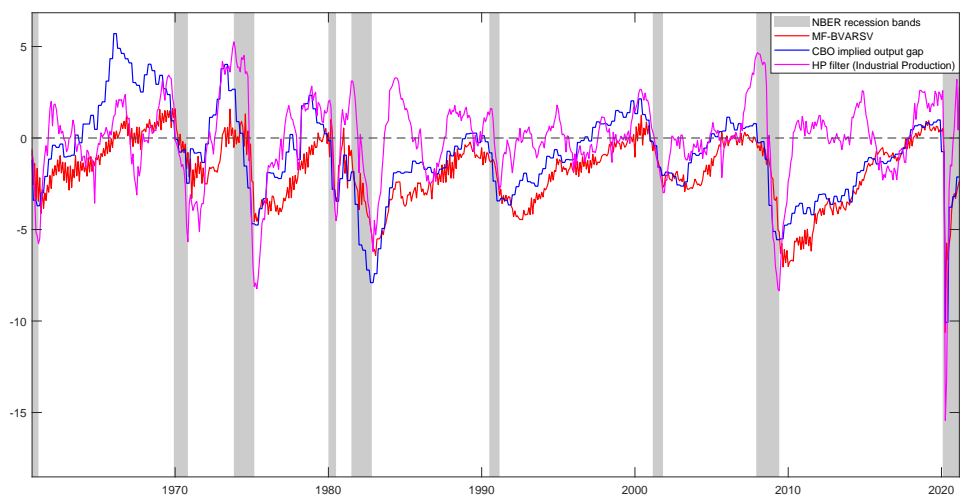}
\caption{A comparison of monthly output gap estimates. 
Notes: The red line depicts the posterior median output gap estimates (BN cycle estimates) from our MF-BVAR-SV model. The blue line represents quarterly output gap estimates derived from the CBO. The pink line depicts an output gap measure derived using the HP filter using monthly industrial production. The grey shaded areas denote NBER recession dates.
}
\label{fig:outputgap}
\end{figure}

Figure~\ref{fig:outputgap} plots the posterior medians of the cyclical component for the monthly real GDP from the MF-BVAR-SV model. These estimates can be interpreted as a measure of the monthly output gap. For comparison, we also plot two alternative measures of the output gap: the CBO implied output gap and an output gap measure derived using the HP filter obtained from monthly industrial production. Since the CBO implied output gap is a quarterly measure, we assume it is constant across the months for each quarter to produce the figure.

It is clear from the figure that the measure derived using the HP filter tends to produce large positive output gap estimates at the start of recessions. In contrast, both our monthly output gap estimates and the CBO measure tend to track the NBER recession dates well. However, there are clear differences between them in specific periods. For instance, from mid-2010 to 2019, our output gap estimates tend to be lower than the CBO measure. 

A potential explanation for these differences could be the inclusion of financial variables in constructing our output gap measure. According to \citet{CGU18}, the CBO uses the production function approach to estimate the potential output. In their modelling approach, they only consider five sectors of the economy, and the financial sector is not included. Therefore, the CBO could potentially underestimate the output gap relative to our estimates after the Great Recession since they do not consider financial variables in their model. In addition, our monthly measure provides more timely output gap estimates for policymakers to monitor the economy in real time. This highlights the importance of incorporating higher frequency indicators in the model when deriving an output gap measure.

\subsection{A Bayesian Proxy State-Space Mixed-Frequency VAR}

In this second application, we show how the proposed mixed-frequency precision-based
samplers can be used for structural analysis. Specifically, we extend the Bayesian Proxy (BP)-VAR in \citet{CH19} to a mixed-frequency setting, which we denote as MF-BP-VAR. While \citet{CH19} estimate both a 4- and 5-equation BP-VAR, here we only consider the endogenous variables from the 5-equation model as that is their preferred model. The five-equation BP-VAR model consists of the
federal funds rate (FFR), the log of manufacturing industrial production
(IP), the unemployment rate (UE), the log of the producer price index
(PPI), and a measure of a corporate spread (Baa corporate bond yield
relative to the ten-year treasury constant maturity), which they denote
as BAA spread. All five of these variables are observed at the monthly
frequency. 

We extend this model to a state-space mixed-frequency VAR by including the log of real GDP, observed at the quarterly frequency. Intuitively, real GDP should be a better measure of the real economy than IP. In
fact, the share of industrial production to real GDP has been declining since the early 2000s. 
For instance, the share of industrial production to real GDP was about 71 per cent at the end of 2000, and by the end of 2007, this share had fallen to about 65 per cent\footnote{These numbers are based on the Real Gross Domestic Product/Industrial Production: Total Index gathered from the US Fred database https://fred.stlouisfed.org/graph/?g=1cnN.}.

We consider two MF-BP-VARs with differernt dimensions. In the first case, we
estimate a 5-equation MF-BP-VAR without IP; it contains 
FFR, UE, PPI, BAA spread, and real GDP. For the second case, we consider all 6 variables, including IP \footnote{It takes approximately 20 minutes to estimate the 6-equation MF-BP-VAR with 100,000 MCMC draws using the standard desktop as specified above.}. We preserve all the model assumptions as specified in \citet{CH19}. We estimate the MF-BP-VARs with $p=12$ lags using data from 1994M1-2007M6 (and 1994Q1-2007Q2 for real GDP). Given that the quarterly real GDP variable enters the model in log-level, we impose an inter-temporal constraint similar to that in \citet{SS15}:
\[
	y_{t}^{GDP,Q}=\frac{1}{3}(y_{t}^{GDP,m}+y_{t-1}^{GDP,m}+y_{t-2}^{GDP,m}),
\]
where $y_{t}^{GDP,Q}$ and $y_{t}^{GDP,m}$ are the observed log quarterly
and latent monthly real GDP variable at time $t$, respectively.

\begin{figure}[H]
\begin{tabular}{c}
\textbf{(a)}\tabularnewline
\includegraphics[width=1\textwidth]{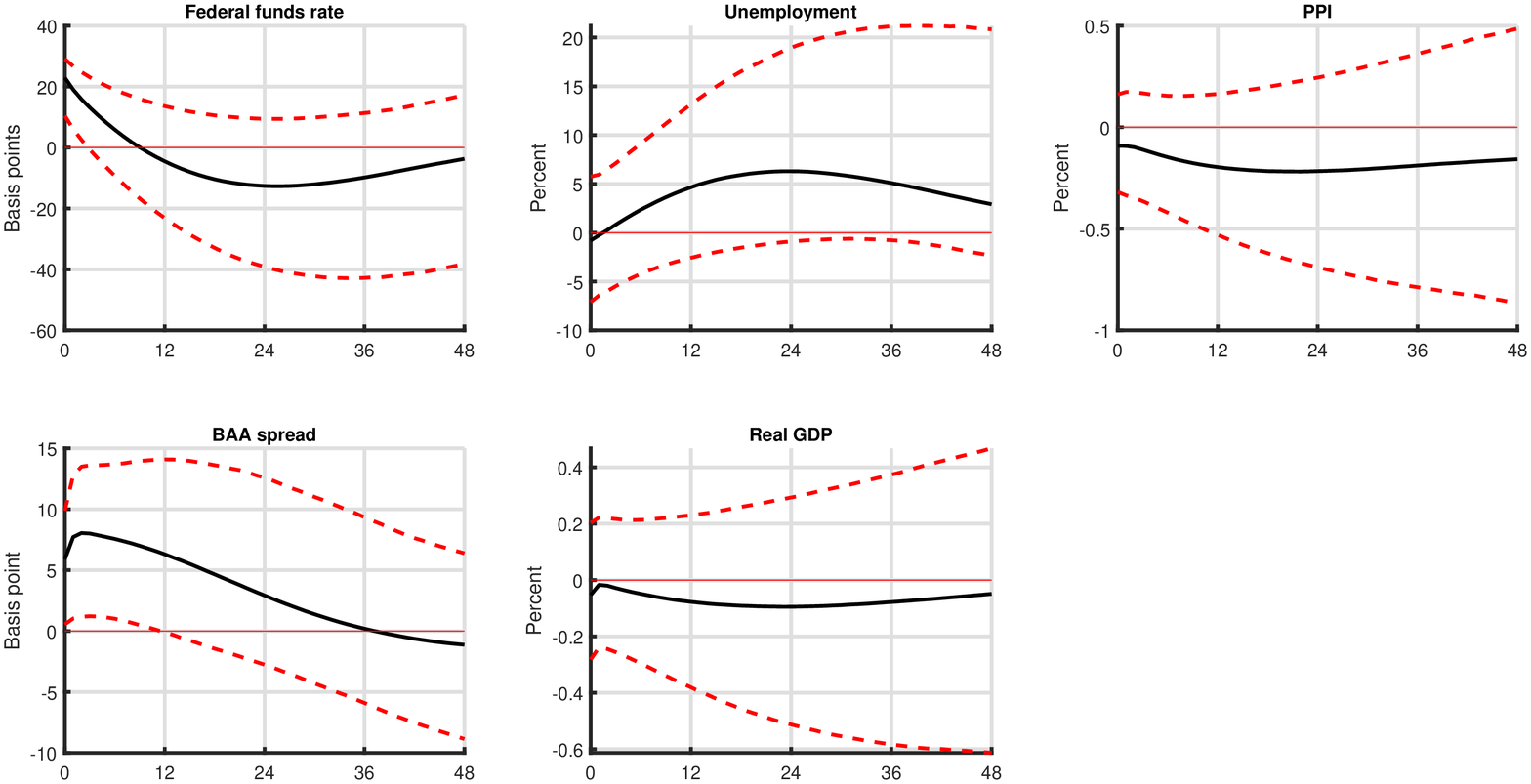}\tabularnewline
\textbf{(b)}\tabularnewline
\includegraphics[width=1\textwidth]{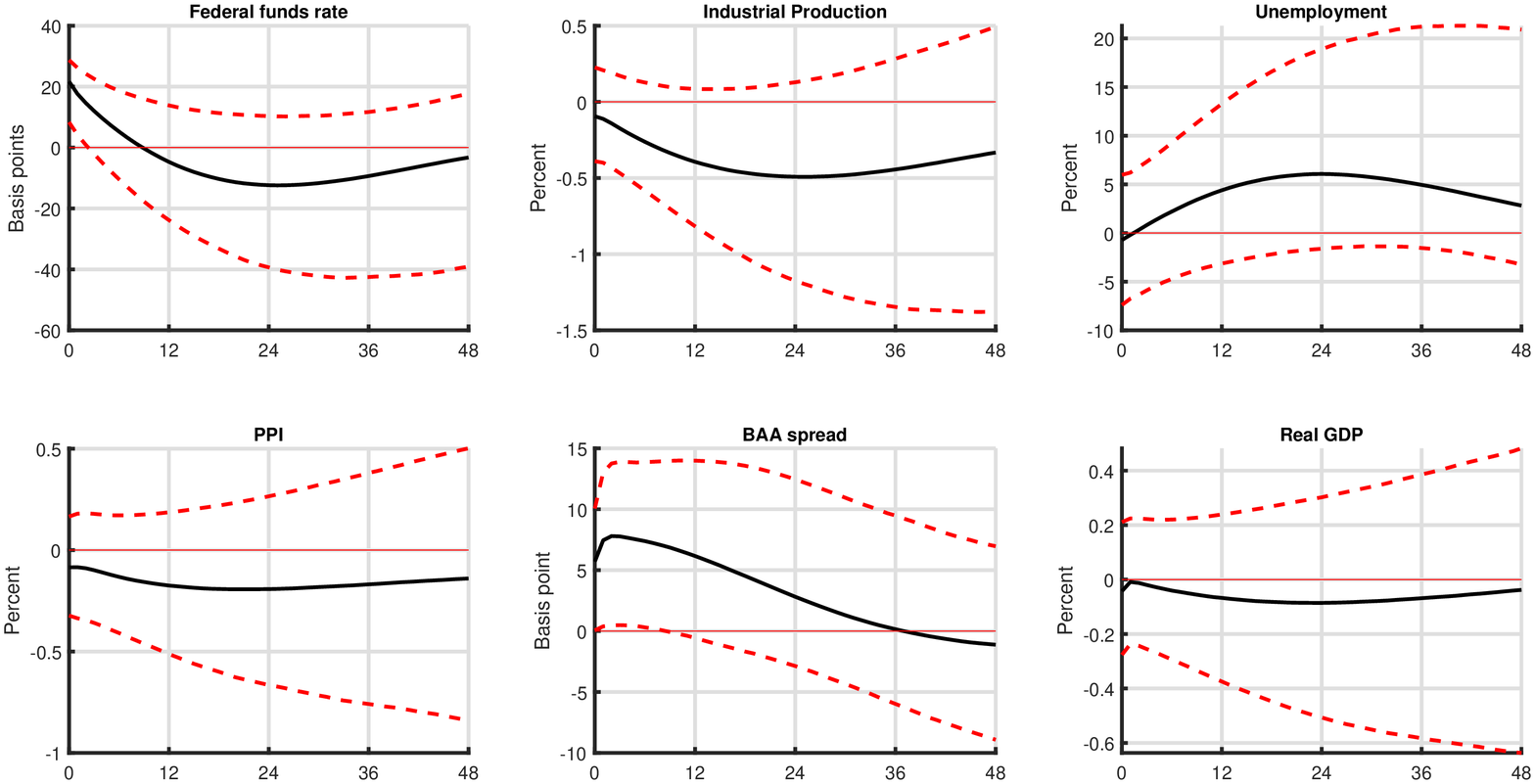}\tabularnewline
\end{tabular}
\caption{Impulse responses to a monetary policy shock. 
Notes: Panel (a) reports the impulse response functions generated from
the 5-equation MF-BP-VAR. Panel (b) reports the impulse response
functions generated from the 6-equation MF-BP-VAR. The solid
black line represents the posterior median impulse responses of the
specified variable to a one standard deviation monetary policy shock
identified via a proxy. The red lines denote the 90 per cent credible
intervals.}
\label{fig:IR}
\end{figure}

Panel (a) of Figure~\ref{fig:IR} displays the impulse response of the endogenous variables
to a one standard deviation monetary shock identified using the 5-equation
MF-BP-VAR, and panel (b) displays the corresponding impulse
response from the 6-equation MF-BP-VAR, which includes IP. The proxy variable used to identify the monetary policy shock in both models is precisely the same as specified in \citet{CH19}. 

All impulse responses of the monthly endogenous variables from both models display exactly the same dynamics as presented in \citet{CH19}. This implies that adding quarterly real GDP variable does not change the dynamics of the monthly endogenous variable to a monetary policy shock. However,
in both the models, the response of real GDP to a monetary policy shock appears to be muted or subdued compared to the response of IP. For example, in panel (a), the posterior median response of IP
falls to a low of $-$0.5 per cent. In contrast, the posterior median
response of real GDP tends to be higher than $-$0.2 per cent on average.
Therefore, this suggests that the response of IP may be overstating
the negative impact of a monetary policy shock on the real economy. 

Table~\ref{tab:coeff} reports the contemporaneous elasticities of FFR to the non-policy variables in both the models and the cumulative elasticities of FFR to all the variables in both models. The majority
of the estimates for the monthly endogenous variables are broadly similar to the results presented in \citet{CH19}. 

\begin{table}[H]
\centering
\caption{Coefficients in the monetary policy equation.}
\label{tab:coeff}
\resizebox{\textwidth}{!}{\begin{tabular}{lcccccc}
\hline\hline
 & \multicolumn{3}{c}{5-equation MF-BP-VAR} & \multicolumn{3}{c}{6-equation MF-BP-VAR} \\
 & Median & 5th \%-tile & 95th \%-tile & Median & 5th \%-tile & 95th \%-tile \\ \hline
\multicolumn{7}{c}{Contemporaneous elasticities}\\ \hline
Industrial production &  &  &  & \textbf{0.07} & -0.25 & 0.47 \\
\rowcolor{lightgray}
Unemployment & \textbf{0.14} & -1.15 & 1.71 & \textbf{0.20} & -1.29 & 1.99\\
PPI inflation& \textbf{0.11} & -0.21 & 0.52 & \textbf{0.10} & -0.25 & 0.60\\
\rowcolor{lightgray}
BAA spread   & \textbf{-1.44}& -4.32 &-0.16 & \textbf{-1.43}& -4.88 &-0.01\\
Real GDP & \textbf{-0.04} & -0.42 & 0.34 & \textbf{-0.05} & -0.48 & 0.37\\ \hline
\multicolumn{7}{c}{Cumulative elasticities} \\ \hline
Federal funds rate & \textbf{0.96} & 0.90 & 1.03 & \textbf{0.96} & 0.90 & 1.04\\
\rowcolor{lightgray}
Industrial production &  &  &  & \textbf{0.00} & 0.00 & 0.00\\
Unemployment & \textbf{-0.05} & -0.18 & 0.06  & \textbf{-0.06} & -0.22 & 0.08\\
\rowcolor{lightgray}
PPI inflation& \textbf{0.00}  & 0.00  & 0.00  & \textbf{0.00}  & 0.00  & 0.00\\
BAA spread   & \textbf{-0.24} & -0.43 & -0.09 & \textbf{-0.21} & -0.43 & 0.00\\
\rowcolor{lightgray}
Real GDP     & \textbf{0.00}  & 0.00  & 0.00 & \textbf{0.00} & 0.00 & 0.00\\ \hline\hline
\end{tabular}
}
{\raggedright \footnotesize{
Notes: The bold entries denote the posterior medians of the contemporaneous elasticities and the cumulative elasticities in the monetary equation identified in the 5-equation
and the 6-equation MF-BP-VARs. The associated 5th and 95th percentiles from the posterior distributions are also reported.}
 \par}
\end{table}

The main results we would like to highlight are the contemporaneous
and cumulative responses to real GDP. The cumulative response of
real GDP for both models is zero. This is consistent with the classical
dichotomy theory, where nominal shocks do not affect the real economy
in the long run. The contemporaneous responses to
real GDP $-$0.04 and $-$0.05 for the 5- and 6-equation MF-BP-VARs, respectively.
This suggests that a one standard deviation surprise
in real GDP, holding all other things constant, will elicit an immediate
monetary policy response of about five basis points. However, this
same interpretation cannot be made for IP as estimates imply 
an inverse relationship between the FFR and IP. 
Furthermore, this highlights the importance of including
real GDP rather than IP as a measure of the real economy.

\section{Conclusion}
We have introduced two precision-based samplers with hard and soft inter-temporal constraints to draw the low-frequency variables' missing values within a state-space MF-VAR. The simulation study shows that the proposed methods are more accurate and computationally efficient in estimating the low-frequency variables' missing values than standard Kalman-filter based methods. We also show how the mixed-frequency precision-based samplers can be applied to two popular
empirical macroeconomic applications. Both empirical applications illustrate the importance of incorporating high-frequency indicators in macroeconomic analysis.

For future research, it would be useful to extend the proposed methods to handle real-time datasets with ragged edges. In addition, developing precision-based samplers for dynamic factor models with complex missing data patterns would be another interesting future research direction.

\newpage

\section*{Appendix: Data}

This appendix provides details of the 22-variable dataset in the first empirical application. Specifically, Table~\ref{tab:data} describes the 22 variables and their transformations. The dataset is sourced from the US FRED database, and the sample covers from 1960M1-2021M3 (and 1960Q1-2021Q1).
All the data is transformed to stationarity. 

\begin{table}[H]
\caption{The list of variables and the corresponding transformation used in the empirical application.}
\label{tab:data}
\resizebox{\textwidth}{!}{\begin{tabular}{lccc}
\hline\hline
Variable & Fred Mnemonic & Transformation & Frequency \\ \hline
Real Personal Consumption Expenditures & DPCERA3M086SBEA & 100$\Delta$log & Monthly \\
\rowcolor{lightgray}
Real Personal Income & RPI & 100$\Delta$log & Monthly \\
Industrial Production & INDPRO & 100$\Delta$log & Monthly\\
\rowcolor{lightgray}
Capacity Utilization: Manufacturing & CUMFNS & $\Delta$ & Monthly\\
Civilian Employment & CE16OV & 100$\Delta$log & Monthly\\
\rowcolor{lightgray}
Civilian Unemployment Rate & UNRATE & Level & Monthly\\
Average Weekly Hours: Manufacturing & AWHMAN & 100$\Delta$log & Monthly\\
\rowcolor{lightgray}
Housing Starts: Total New Privately Owned & HOUSTS & 100$\Delta$log & Monthly\\
Personal Consumption Expenditures: Chain Index & PCEPI & 100$\Delta^{2}$log &Monthly\\
\rowcolor{lightgray}
Consumer Price Index: All Items & CPIAUCSL & 100$\Delta$log & Monthly \\
PPI: Metals and metal products: & PPICMM & 100$\Delta$log & Monthly\\
\rowcolor{lightgray}
PPI: Crude Materials & WPSID61 & 100$\Delta$log & Monthly\\
Avg Hourly Earnings: Manufacturing & CES3000000008 & 100$\Delta^{2}$log & Monthly\\
\rowcolor{lightgray}
Real M2 Money Stock & M2REAL & 100$\Delta$log & Monthly\\
Total Reserves of Depository Institutions & TOTRESNS & 100$\Delta$log & Monthly\\
\rowcolor{lightgray}
Commercial and Industrial Loans & BUSLOANS & 100$\Delta$log & Monthly\\
Effective Federal Funds Rate & FEDFUNDS & $\Delta$ & Monthly\\
\rowcolor{lightgray}
10-Year Treasury Rate minus 1-Year Treasury Rate & GS10 - GS1 & Level & Monthly\\
\rowcolor{lightgray}
Moody's Seasoned Baa Corporate Bond Yield & & & \\
minus Moody's Seasoned Aaa Corporate Bond Yield & BAA-AAA & Level &Monthly\\
\rowcolor{lightgray}
U.S. / U.K. Foreign Exchange Rate & EXUSUKx & 100$\Delta$log & Monthly\\
S\&P's Common Stock Price Index: Composite &S\&P 500 & 100$\Delta$log &Monthly\\
\rowcolor{lightgray}
Real Gross Domestic Product & GDPC1 &100$\Delta$log & Quarterly\\ \hline\hline
\end{tabular}
}
\end{table}

%
%

\newpage

\singlespace

\ifx\undefined\BySame
\newcommand{\BySame}{\leavevmode\rule[.5ex]{3em}{.5pt}\ }
\fi
\ifx\undefined\textsc
\newcommand{\textsc}[1]{{\sc #1}}
\newcommand{\emph}[1]{{\em #1\/}}
\let\tmpsmall\small
\renewcommand{\small}{\tmpsmall\sc}
\fi

\end{document}